\documentclass{article}
\usepackage[T1]{fontenc}
\usepackage{lmodern}
\usepackage{graphicx}
\usepackage{epsfig}
\usepackage{bbm}

\begin{document}

\title{The battle of High Temperature Superconductivity }

\author{P. Lederer \\ 
Directeur de Recherche honoraire au C.N.R.S.\\
14 rue du Cardinal Lemoine, 75005-Paris\\
e-mail: pascal.lederer@u-psud.fr\\
33(0)662984051 }

\maketitle
\newpage

\begin{abstract}
The early development of conflicting theories  about the microscopic mechanism of High Temperature Superconductivity is described. The biographical roots of this diversity are stressed, as well as its subjective/objective roots. This study of a specific case of knowledge about a specific fact of nature allows to discuss the subjective and objective roots of scientific pluralism.   Relativism, the Duhem-Quine thesis on the underdetermination of theory by facts, are discussed from the stand point of the materialist view on the  dialectics of knowledge and nature. Developments of dialectic materialism seem to be suggested by this study.

\end{abstract}

\vspace{20mm}
\section{Introduction}
This paper is interested in discussing a current puzzling situation in contemporary physics: a plurality of conflicting theories have emerged very rapidly upon the discovery, in 1986,  of High Temperature Superconductivity  ( hereafter HTS). Understanding how,  why, and if  HTS is a new phenomenon has led to a variety of theoretical proposals, which  have not stopped since then to confront each other.

 My main point of view here is that of dialectic materialism, a point of view which is rejected with various nuances of outrage or contempt by a large current of analytic philosophers nowadays. In fact, I wish to point out that, in my view, the battle about HTS suggests that qualitative enrichments of dialectics are needed : in general the subjective/objective approach of dialectical materialism to the connection between knowledge and reality relies on the analysis of possible coexistence and even identity of bipolar contraries within the thing: cause and effect, quality and quantity, capital and work,  corpuscular and wave like behaviour, kinetic energy vs potential energy, etc.. What the study of HTS suggests, as I will argue,  is that a multipolar set of contradictions emerges from experiments and theories, a suggestion which might help overcome the present situation in HTS of conflicting theories which ignore or fiercely fight each others.

Given the degree of present day lack of interest for, or misconceptions    about dialectical materialism among many philosophers of science, I deem it necessary to devote to this topic a whole section (section \ref{contradiction}), before entering the topic of HTS proper,  in the hope that some readers will admit reading the paper to the end to judge its value in an objective (however also necessarily subjective) way. That section, for want of space, is admittedly elementary,  schematic and incomplete. I am inspired by  Lucien S\`eve's point of view, as developed (in French) in  a series of recent books \cite{seve}. The interested reader should rely on those works for a detailed and more convincing defence of the dialectical materialist point of view.

Section \ref{pluralism} is a brief account of contemporary discussions on scientific pluralism.

Section \ref{HTS} sets the stage of HTS: how it started in 1986, why it was a total surprise, what are the major features which are agreed upon by most physicists as far as a vast class of HTS materials is concerned; the contrast with the previously known and seemingly well understood phenomenon
of superconductivity since 1911 is underlined.

Section \ref{history} is an outline of the stunning development of contemporary controversies, since 1986, among physicists about the theory of HTS. This outline,  as honest as I have tried to write it down,  is necessarily both  subjective and objective. This is because I have been personally involved in the HTS research, a fact which necessarily gives me both a  somewhat learned but subjective/objective point of view.  Given the incredible volume of experimental and theoretical publications (probably more than hundred thousands in tens or hundreds of scientific journals), no one at this stage is able to give a complete balanced account of HTS. This suggests that a collective effort of physicists and philosophers of all trends, to analyze the HTS phenomenon from a philosophical point of view, would be necessary and  welcome. Knowledge of superconductivity, and especially about HTS, among philosophers of physics is, unfortunately, not as general as it should be; this paper hopes to stimulate bridging this gap in philosophical discussions.

That section is somewhat technical; it requires some elementary knowledge of condensed matter physics. I hope it remains nevertheless useful for the readers less familiar with this area of physics.

Section \ref{underdetermination} is devoted to a recurrent theme in epistemological discussions: the Duhem-Quine thesis on the ``underdetermination of theory by the facts''. HTS may seem to be a good example of this thesis, which is at the center of neo-kantian positions, following which the thing in itself (das Ding in sich) escapes for ever a complete knowledge. I argue in that section that this thesis does not have universal value and cannot be the final word about HTS.

\section{On the limits of the no-contradiction principle}\label{contradiction}

This section  is devoted to the question of the ``contradiction within objects of nature'' which I have mentionned in the introduction. There is currently a large sector among philosophers and in particular philosophers  of science who consider the ``no contradiction principle'' established by Aristotle has such a validity that any mention of coexistence and - worse, identity - of contraries in objects of nature is immediately rejected by many, with no hesitation or discussion. Some \cite{kojeve}  admit contradictions within social objects, such as the  capital/work antagonism. However admitting contradictions within nature is rejected. This last position  neglects the fact that humans,  and their society, are ultimately  also objects of nature, except from a spiritualist or idealist point of view. If contradictions within a material object of nature such as human society is admitted, there is no rational argument to ban them within other objects of nature.  Even though there are very good reasons for holding to the no-contradiction principle,  there are also good reasons to reconsider it, and its limits of validity,  in view of some arguments I discuss now.

The history of  dialectical thinking is about three thousand years old \cite{seve}.That some truth may be expressed in a contradiction is at the heart of the most ancient philosophies in China, India, Greece, from Lao-tseu to Heraklite \footnote{Example from Heraklite: `` It is impossible to bathe twice in the same river''}. This was radically dismissed by Aristotle \footnote{Aristotle : ``the same cannot belong and simultaneously not belong to the same simultaneously and under the same connexion''(ontology); ``contradictory statements cannot be simultaneously true'' (logic); ``nobody can believe that the same could be simultaneously be and not be''(psychology)}, in the name of a logic he established \cite{aristo}.  Dialectical thinking was  banished in the western world   as detrimental to social order, also as  making  communication impossible.

 A crucial point is that the no-contradiction principle is based on an ontological postulate, i.e. the invariance of the essence. Masked by the universal sensible change  in our sublunar world, the latter is  nevertheless  the ultimate truth of the being\cite{seve,aristo}. 

Then Kant, at the time of the French Revolution \cite{kant} finds out that the century old efforts of metaphysics results in fundamental and unsolvable contradictions, which he dubbed antinomies. The latter, says Kant, are a sign of a fundamental incapacity of our understanding. 

Then Hegel \cite{hegel} takes a bold new stand: if all quest for truth inevitably results in a contradiction, then contradiction is the truth!

For Hegel, dialectics is not illogical, it is logic developed beyond the limits of aristotelian logic. The no-contradiction principle appears to be relevant for the invariant, the  inert thing, but it is in great difficulty to think connections and processes, as illustrated by the problems posed since ancient times by the simple motion of an arrow...Hegel's ideas caused enormous interest in Europe in the 19th century, as well as fierce opposition, in particular by the catholic church, for which Hegel is a pantheist, or worse, an atheist\footnote{I cannot review here those developments, for want of space. Interested readers are invited to read, for example, the recent book by S\'eve (chapter IV) \cite{seve}.}. The 2nd Empire and Napoleon the 3rd banish Hegel's ideas from Academia. In turn the working class movement and the rise of socialist thinking  triggered  a revival of hegelian studies. Marx \cite{marx} had adopted hegelian dialectics, but rejects the idealist hegelian position following which the Idea, the Concept, precede reality; for him, dialectics appears in the theory of knowledge as a result of dialectics in the objective world. Engels \cite{engels1} develops dialectical materialism, formulates laws thereof; at first dialectics is for him objective inasmuch as it is imposed by reality to our subjective logic; later  he reaches a disputable position: dialectical materialism becomes, in a  pure ontological way  {\it{the science of the general laws of motion of the external world as well as those of human thought}}\cite{engels2}: this thesis dismisses the essential epistemic aspect of dialectics and eventually opens the way to the catastrophic version of stalinist dogmatism.

After the defeat of nazism, dialectical materialism became an important philosophical current. Most serious philosophers in Europe, Asia and elsewhere adopted it, in one version or another. This success led to its demise. It suffered a severe blow when it was used as official state philosophy in the USSR.  Much to  the contrary, nothing,  in the founding philosophical writings \cite{marx,engels1,lenin} allowed to justify turning them into an official State philosophy.  This produced  such catastrophies as  the State support for Lyssenko's theories, based on the notion that genetics was a bourgeois science, while lamarckian concepts were defined at the government level as correct from the point of view of a caricature of dialectical materialism. It is understandable that such nonsense in the name of a philosophical principle turned the latter into a questionable construction in the eyes of many.

Dialectic materialism itself is an open system, which has no lesson to teach beforehand about specific objects of knowledge, and insists \cite{engels1} on taking into account all lessons taught by the advancement of science. This is precisely my outlook in this paper: the story of High Temperature Superconductivity may indicate that dialectical thinking about nature needs some conceptual development.

 It is perhaps time for a serious critical assessment of this philosophical current and of its principles. The possibility of general theoretical statements  about the empirical world is not a negligible question. 

 Materialism gives a clear answer to the "`{\it{gnoseological problem of the relationship between thought and existence, between sense-data and the world...Matter is that which, acting on our senses, produces sensations.}}'' This was written in 1908 by Lenin \cite{lenin}. It may look too simple when  technology (such as that used in quantum physics  experiments) is intercalated between matter (for example the  electrons in the two slit experiment and the screens on which we read their impact).  Technology or not, matter is the external source of our sensations. So much for materialism. Dialectical materialism adds  a fundamental aspect  i.e. that theories reflect the existence of  contraries which  coexist and compete with each other within things in Nature. Depending on which dominates the competition (contradiction) under what conditions, the causal chains originating from the thing and causing phenomena will take different forms, which are reflected in theories. Epistemics and ontology are intimately intertwined. 

Consider an example of how formal logic and dialectical logic  complete and enrich each other: that of cause and effect. A moving billiard ball 1 hits a motionless billiard ball 2 which is thus set in motion. The motion of 1 is the cause for the motion of 2, which is the effect. Cause and effect are two contraries of a logic of identity: their meaning is clear, the relation is uni-directional: there seems to have no room left for contradiction. However, if the collision has caused the motion of 2, due to that of 1, the trajectory, energy  and  velocity of 1 have also been changed; to the initial causal relationship wherein 1 is a cause for 2 is added necessarily an inverse causal effect wherein 2 becomes  a cause for the motion of 1. The uni-directional causal relation we had first is turned immediately in a reciprocal relation: a cause leading to a consequence is in turn affected by the consequence turned into a cause itself. Can't we see here an example of unity of contraries? The classical logician will deny it, observing that there is no contradiction, but interaction: coincidence of two causal actions which remain distinct causal ones. However, how can we distinguish within the collision the causal action of 1 on 2 and that of 2 on 1?  

Another example is given by Aristotle's fundamental categories such as quality and quantity. In the classical logic, those two categories are clearly distinct and form a couple of well identified contraries. But there are  hosts of empirical evidence that there is no such  dichotomy : quantity transforms into quality almost universally.

\subsection{Dialectics of knowledge, dialectics of nature, dialectical materialism}\label{dialectics1}

	A central philosophical problem is that of the relationship between thought and being. Following S\`eve \cite{seve-intro} \textit{...this defines  a central structure  of the network of philosophical categories. The relationship between being and thought is indeed dedoubled  in the \underline{thought about the being}, and the \underline{being of the thought}, i.e. in the categories of the objective reality  reflected  by  thought, and those of the subjective reflection of the being. This relationship is not abstractly motionless, it is a process, which is itself a double one: the motion of the being and the motion of thought. From the latter, four groups of categories emerge: 
	\begin{itemize}
	\item (thought of) the being in motion: materialist outlook on the world,
	\item (being of) the thought in motion: materialist outlook on knowledge,
	\item (thought of) the motion of the being: objective dialectics,
	\item (being of) the motion of thought: subjective dialectics
	\end{itemize} 
	Matter, contradictions, reflection, concreteness: such are the central categories of those four groups : the concrete reflection in thought of the contradictions of matter : such is the essence of the relationship of thought to being. But it is crucial not to forget the dimension of the practice; as emphasized by Engels, ``the practical activity of man in nature -- not only nature as such -- is the most essential cause, the most direct one, of human thought''}. Note than ore than hundred years after Marx, Hacking \cite{hacking} rediscovers practice as a criterion of reality, without, nevertheless, adopting a materialist view of reality: ``realism'' does not explicitly bans God from the possible real entities.
	
	S\`eve \cite{seve} writes also: \textit{...the two poles of the cognitive process, the subjective and the objective must be recognized not only as opposite ones, but as identical: the objective part is subjective -- and here the very meaning of the scientific endeavour seems to be lost -- but simultaneously the conditions and limits  under which the subjective part is objective become determinable.}
	
	Popper \cite{popper} in his book \textit{Objective Knowledge} considers the world of ``\textit{objective contents of thought}'', but considers that contradictions have to be eliminated. Although is thesis is an evolutionary approach, he misses the richness of the four poles of categories mentionned above, and dismisses the value of Hegel's philosophy.
	
	What will be described below about HTS illustrates the relevance of the four points above.

\section{Scientific pluralism}\label{pluralism}

What is at stake in a number of discussions about conflicting theories may be interpreted as the question of ``scientific pluralism": can it be that two or more conflicting theories account for the same phenomena, or account  for  the causal powers of the same objective real entities?

Dickson \cite{dickson} defines scientific pluralism as "`{\it{...the existence or toleration of a diversity  of theories, interpretations, or methodologies within science}}"'.

 Following Hillary Putnam \cite{putnam}, who discusses how things (e.g. electrons) are named, and how theories evolve, all sorts of incompatible accounts of the thing appear, all of which agree in describing various causal powers which may be employed while acting on nature. 

Cartwright \cite{cartwright} emphasizes that in several branches of quantum mechanics, searchers may use  a number of different models of the same phenomena. They can be mutually inconsistent, and none is the whole truth.

Dickson asks: {\it{"`How can one be a pluralist about science while respecting the (approximate) validity of our best scientific theories?"'}} In the course of his paper, he defends his own admission that as far as quantum mechanics is concerned, pluralism is justified and acknowledges the existence, and toleration of  a diversity of contradictory theories. He writes: {\it{I shall address  the most obvious and serious objection to such a view...namely, that it places the scientifically minded person in the intolerable position of explicitly endorsing contradictions within science (as a matter of principle and not merely as a pragmatic matter). To do so is to reject the scientific enterprise.}}

The authors in references \cite{dickson,putnam,cartwright,hacking} admit that there might be pluralism in theories, in various guises. But admitting  ontological contradictions within the thing seems to all authors, except perhaps Putnam, to be  prohibited  and intolerable. All adhere to the no-contradiction principle as an unquestionable one.

A straightforward way out of the problem of scientific pluralism is to adopt one version or another of relativism. Following
 Kuhn\cite{Kuhn} in {\it{The Structure of Scientific Revolutions}}, ``scientific theories are beliefs shared by groups of persons''\footnote{Kuhn himself has been critical  on the relativistic interpretation of his views.}. If such is the case, if theories are merely social constructions, as other authors contend \cite{rorty,latour}, contradictions between theories and battles between theorists can be accounted for by social, cultural, historical differences between ``groups of persons'': they do not signal anything about ontological contradictions.

Dickson describes various sorts of pluralisms (it seems one may define 27 types).  What I have been defending in a previous paper \cite{lederer} is that contradictions in epistemics may reflect  disagreements between erroneous vs correct theories --admittedly a rather trivial sort of pluralism -- , etc., but may also reflect\underline{ ontological contradictions}. As discussed in the previous section the  aristotelian prohibition has its domain of validity, and its limits of validity. One has to face the possibility that accepting contradictions in nature is a way to save the scientific enterprise when science is pluralist. 

I do not agree with most things Dickson says about quantum mechanics\footnote{Such as stating that quantum mechanics has no dynamics. I cannot enter in details here about these disagreements.}.
 I stress here that he, a philosopher convinced of the aristotelian prohibition of contradictions, is constrained to admit them  as a true feature of  theory.

Following  section \ref{contradiction}, the scientific endeavour is itself  rich with contradictions: it is a social process, through which individuals and groups confront their subjective/objective views, while theorists are increasingly separated from experimentalists; the latter use increasingly sophisticated instruments which access increasingly varied properties of matter, or phenomena,  with increasing accuracy in all manners of measurements. If one accepts that the evolution of matter is dominated by the struggle of contraries which coexist within the thing, the emergence of a true theory is almost necessarily a socially contradictory process: scientific pluralism has to be a fairly general stage of science, leading in time  to more complete theories which overcome the previous controversies in  which single physicists, or groups, had taken  part, until errors have been eliminated, and ontological contradictions are properly taken into account as part of reality.


My main disagreement with all neo-kantian philosophers quoted above is that scientific pluralism is an almost  unavoidable but  transient feature of the scientific development.  For Newton, light is corpuscular. For Fresnel, it is a wave. It takes two centuries of controversies and seemingly contradictory experimental results to understand that those contradictions in epistemics reflect the quantum mechanical behaviour of microscopic systems. Both views retained a fundamental part of the truth, while the aristotelian no-contradiction principle implied that one or the other was wrong. 
	
	This particular  example has universal value: contradictions in epistemics  may well reflect, at times, ontological contradictions.
	Phenomenal manifestations of the same object of nature differ when one pole of an ontological contradiction dominates, or the reverse. The conditions for one 
	pole to dominate on the other depend on the various constraints exerted by the environment, such as pressure, temperature, magnetic field intensity, etc.. Analyzing fully the contradictions within the thing is the way to reach scientific truth. Is this notion of interest in the problem of High Temperature Superconductivity? This question is central to this paper.

\section{High Temperature Superconductivity} \label{HTS}

The history, physics and theory of Superconductivity are rich with surprises, suspense, and, in the last 30 years, with fierce scientific battles among theorists and experimental groups. Never before in the history of science have so many papers been published about a specific topic  with no agreement in sight about the theoretical understanding of the microscopic mechanism of what appeared as a coup de th\'e\^atre in 1986: the astonishing discovery by an aging and obscure scientist and his assistant\footnote{Both were jointly awarded a Nobel Prize...} of superconductivity, at temperatures much higher than ever  reached before for this phenomenon, in  compounds  where no educated physicist, except perhaps one or two\footnote{See however reference \cite{chakra}}, would ever had bet one cent on its appearance \cite{bednorz}.

As I had started writing this paper, I stumbled upon a review article by a physicist on the the same topic \cite{zaanen}. The author writes:\begin{quote} {\it{ One cannot write the history of a war when it is still raging and this is certainly applying to the
task we are facing dealing with the theory of superconductivity in the post BCS era \footnote{The BCS theory was the  recognized theory for the vast majority of superconductors known before 1986}. The BCS theory was of course a monumental achievement that deserves to be counted among the greatest
triumphs in physics of the twentieth century. With the discovery of high-Tc superconductivity in the cuprates in 1986 a consensus emerged immediately that something else was at work than the classic (i.e. phonon driven) BCS mechanism.}}\end{quote}

Superconductivity first  erupted in the world of scientific knowledge in 1911 -- long before  what is the  topic of this paper --  as a consequence of a scientific and  technological advance. Kammerlingh  Onnes \cite{KO}, in 1908, managed to liquidify Helium at the incredibly  -- at the time -- low temperature of about 4 degrees Kelvin \footnote{Nowadays there are ways to reach temperatures of  a millionth of a degree...}. This  allowed to explore the properties of matter within new, extended  ranges of parameters. 
With this achievement, Kammerlingh Onnes and his collaborators could endeavour to  study the behaviour of condensed matter at hitherto unexplored low temperatures. Physics at the time was very much geared to exploring properties of metals and alloys, because of fundamental reasons, and also because of the industrial interest in such bodies, in particular because of their electrical and thermal conducting  properties. It was already known that down to liquid nitrogen temperatures (around 70 degrees K) all metallic resistivities were decreasing functions of temperature. Various laws describing this decrease had been observed in various metals. When the laboratory in Leyden studied the resistive behaviour of Mercury (noted Hg), a mysterious abrupt drop in resistivity was observed at 4 degres K, and the metal seemed to become a perfect conductor! Various metals were found then to exhibit a similar behaviour, at the so called ``critical temperatures'' $T_c$ which are  characteristic of each pure metal. It took twenty more years to realize that this phenomenon was not simply a perfect conductivity, but a conductivity of a special type: it  expels the magnetic field \cite{MO} from the volume of the metal\footnote{The magnetic field is expelled at sufficiently low fields; the story is more complicated when the magnetic field intensity is larger than a so-called critical field called $ H_{c1} $}. This is not what would  happen in a normal metal if its resistivity became zero\footnote{for example in perfectly pure metal.}. 

The long and difficult quest for a theoretical understanding for this phenomenon is a fascinating story in itself, but it is not the topic of this paper. A major theoretical advance (quoted above as the BCS theory) was achieved by Bardeen Cooper and Schrieffer \cite{BCS} in 1958 when they published a revolutionary theory: superconductivity was due to a breakdown of the normal metal electronic structure: instead of behaving as almost independent particles governed by Fermi statistics, electrons (which have half integer spin and therefore are fermions) interact with lattice vibrations (called phonons) of the metallic crystal. The latter may be viewed as sort of glue which dominates the repulsive Coulomb interaction between electrons at low enough temperatures (incidentally, attractive vs repulsive interactions are yet another example of contradiction within the same thing). As a result electrons become associated in pairs, which are called singlets, have zero spin, and are bosons. Identical bosons had been shown by Bose and Einstein to condense in a superfluid state a low enough temperature. A superfluid state of charged particles is a superconducting state. Ten years of intense exploitation of the BCS theoretical advance followed, somewhat along the lines of what Kuhn called normal science \cite{Kuhn}. New theoretical entities became familiar terms: order parameter, phase coherence, penetration depth, vortex phase, etc.. The order parameter -- in the case of BCS superconductivity a complex number --  characterizes the new superconducting quality: it starts from zero at the superconducting critical temperature when the temperature decreases, and its intensity grows up to a maximum at zero temperature, with thermal variation laws which were determined experimentally and explained theoretically within the BCS theory. Magnetic fields, and magnetism at large, were well identified as detrimental to (contradictory with) superconductivity and as a cause for its destruction. New devices appeared \cite{squid},  based on quantum properties of the superconducting state in the presence of magnetic fields. They are now  standard tools for scientific investigation, or industrial and medical applications.

Some years around 1970, it appeared to the overwhelming majority of physicists and science managers -- public or private --  that the (scientific) gold rush to superconductivity had exhausted its scientific novelty deposits. Science leaders who had been prominent in developing the field dropped it as a research program altogether, more or less abruptly, as miners abandon a gold mine when its yield becomes negligible. Scientific teams were disbanded\footnote{In Japan, a small team of scientists was instructed by the science ministry to  keep watching the physics reviews in case some improbable unforeseen novelty appeared in the field of superconductivity. } and geared to other fields: everything about superconductivity was known. The BCS theory was universally held as true. It had become   {\it{an obsolete field }} \cite{degennes}. The loss of interest was also due to the quite commonly shared view among physicists that the highest possible superconductivity temperature on the planet earth would never exceed $25$ K by more than a few degrees. This rule, which had been formulated empirically  by a well known experimentalist meant that superconductors would remain for ever a laboratory phenomenon, with almost no industrial application. 

Only a handful of searchers, mostly aging and obscure ones, kept trying to find ways of realizing higher superconductivity temperatures. A major industrial  and environmental interest, to this day, is  to find ways of storing electricity in a cheap way: superconducting currents, circulating in a supercondcting ring,  do not suffer the costly energy losses due to Ohm's law for usual electric currents: they last --in principle -- for ever. Finding ways of storing electrical energy as chemical energy is stored in coal, oil, or...dynamite, is a long lasting industrial dream. If this storage process requires  prohibitive costs to cool down conductors to such temperatures as 20 K, the costs exceed the benefits. 

When High Temperature Superconductivity (hereafter HTS) was discovered, this dream suddenly seemed to come true. Whole teams of physicists rushed to study, understand  and improve the totally new and unexpected material -- chemically doped copper oxides -- which  exhibits HTS.

\begin{figure}\label{structure}
\centering
\includegraphics[width=10.5cm,angle=0]{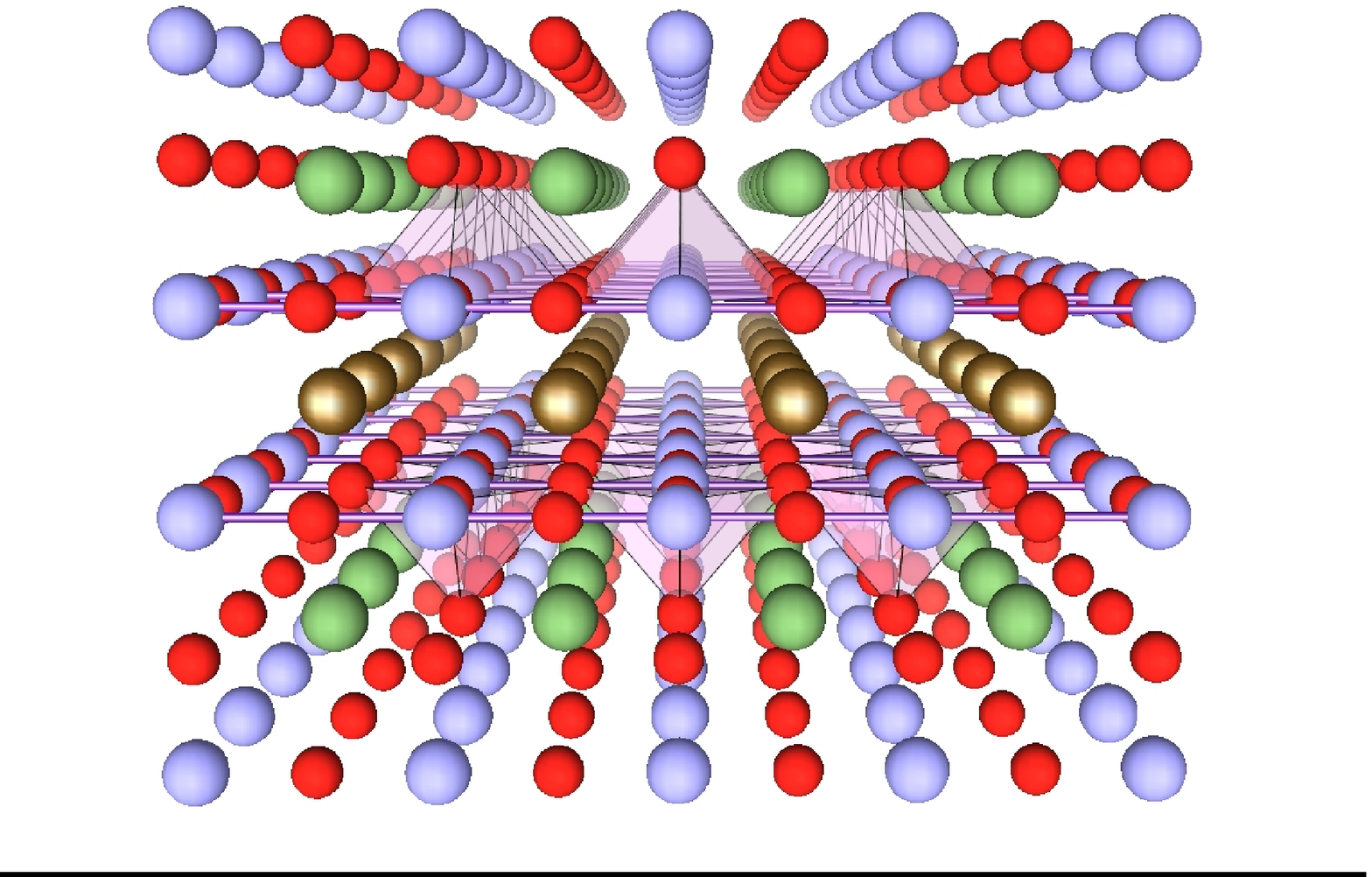}
\caption{\footnotesize{
This figure exhibits the atomic structure of a typical HT Superconductor: contrary to the vast majority of (lower temperature) BCS superconductors, which have a much simpler chemical structure and a much simpler (usually cubic) crystallographic one (in one guise or another)), this crystal is formed by parallel sheets of Cu-O  (Copper oxide) planes. The ``red'' atoms here are oxygen atoms, the ``blue'' ones are Cu (Copper)atoms;  two Cu-O planes  are separated by ``brown'' atoms (such as Ytrium or Lanthanum } while ``green'' atoms are divalent metals such as Ca or Ba. A large family of HT Superconductors exhibits the common feature of weakly coupled Cu-O planes which have a basic square elementary cell. The occurrence of linear arrays of Oxygen chains between copper planes is specific to the  particular  chemical (YBaCuO) shown above. I am indebted to Julien Bobroff for providing me with the file for this figure. }
\label{fig01}
\end{figure}

 Figure \ref{fig01} shows a schematic view of the chemistry and structure of a typical HTS material. One should add to  Figure \ref{fig01} that by altering the stoechiometry of various chemical components, such as Oxygen, or substituting a small concentration of La atom by Sr atoms, or some other chemical manipulation, one can suppress an equivalent concentration of electrons -- i.e. inject ``holes''--. In fact the stoechiometric compound (no hole) is an antiferromagnetic (AF) insulator, which becomes superconducting with increasing superconducting temperature upon increasing the concentration of holes.

 Very quickly the family of doped copper oxides became quite large; later on new families were discovered, such as Fe pnictides. As of today, the largest known superconductivity temperature is that of  a Bismuth Copper oxide compound and is  around $150$ K, i.e. some $80$ K larger than that for liquidify Nitrogen.

Thirty years after HTS was discovered, its fundamental mechanism is still the object of passionate debates among physicists.

The aim of this paper is to explore some philosophical lessons taught and questions raised  by the HTS story until now. Putting aside the ontological question about the reality of HTS, which is sufficiently established by all converging tens of thousands of experiments and industrial applications, the main discussions will be about:
\begin{itemize}
\item the subjective roots of this diversity;
\item the objective (ontological) roots of this diversity;
\item the Duhem-Quine thesis about the ``under determination of theory by experiments'' \cite{duhem,quine}. Is the war mentionned above doomed to last forever?
\item the lessons taught by this story from the point of view of the dialectics of knowledge and/or that of  nature \cite{seve,lederer}.
\end{itemize}

\section{Historical development of different theories of HTS}\label{history}

\subsection{Agreements and disagreements: subjective contradictions }

\begin{itemize}
\item Agreements.

As soon as the experimental paper on the new HTS appeared, a flurry of theoretical differences among the various theories of the phenomenon appeared, on a background of universal theoretical agreement. The agreement was, with no exception known to me, that any theory had to account for the instability of the normal electronic system on the basis of formation of bosonic electron pairs. The HTS displays  zero resistance and the Meissner effect; as for BCS superconductors, HTS is a ``spontaneous breaking of the gauge invariance'' of the normal electron system.  This is the common background for the different theories concerning the microscopic mechanism of HTS.
\item Disagreements.

Disagreements had their roots in the evaluation by the various schools --should I say chapels? -- of the specific   reason why the new material was falsifying the previous wisdom on the formerly known  superconducting BCS material. Some of the disagreements are deduced by inspecting the figure \ref{fig01}. Depending on which structural property of the new material is deemed to be at the basis of the new behaviour, different theoretical proposals will emerge.
The specific Copper oxides structural features, as compared to  ``ancient'' BCS metals are:

1.-- contrary to most BCS metals which have cubic symmetry, HT Superconductors are organized as stacks of CuO planes --with a square array of Cu atoms separated by Oxygen -- in weak electronic contact with each other, as these planes are in general separated from each other by a layer of inert atoms. In fact, differing on the specific HT Superconductor, the structure may have  separated CuO planes, or  sandwiches of two or more  CuO planes   separated from other sandwiches. HTS are anisotropic  crystals with a preferred symmetry axis perpendicular to the CuO planes.

 2.--In a CuO plane, since each Cu carries one outer electron available for conduction, the system of conduction electrons -- if repulsive Coulomb interactions are neglected -- should form a two-dimensional (2D) square Fermi surface occupying half of the available electronic states. This is in contrast with the 3-dimensional (3D) spherical  Fermi Surface considered by the BCS theory.

3.-- The oxide character of the HTS compounds  is quite different from the simple monoatomic structure of most BCS superconductors: the chemical structure is qualitatively more complex.

4.--  An astonishing feature of the CuO HTS is that the isolated CuO plane is  theoretically a textbook example  of an AF insulator: each Cu carries an electron bound to this atom. Two  electrons spins on neighbouring Cu atoms are coupled \footnote {by a coupling mechanism known since the fifties called super-exchange.} antiferromagnetically, which means that the electronic lattice, at zero temperature should be a regular N\'eel  array of localized spins with alternating spin direction from atom to atom, resulting in a so-called ``Mott insulator''. The latter  is based on the notion that Coulomb interactions are so strong between electrons  that they are constrained to stay apart from one another and stay localized each on its Cu atom. The AF order results from a small  energy gain when neighbouring spins are antiparallel.  This is in contrast with item 2 above, where Coulomb interactions are assumed to be  negligible, so that electronic wave functions spread over the whole CuO plane, turning the crystal with one electron per atomic cell in a good metal. Furthermore, the well established experimental fact that HTS appears as emerging upon doping a magnetic material was at variance with the BCS wisdom that magnetism destroys superconductivity.

5.-- In addition doping the insulating crystal to turn it into a HTS inevitably introduces disorder in the lattice.
\end{itemize}

\subsection{The subjective/objective biographical roots of  scientific pluralism in the history of HTS.}\label{biography}

Below, I list some of the main conflicting theoretical proposals which have been formulated very early on, and how each has been rooted in the past scientific activity of its champions. They are ordered in increasing degree of novelty with respect to the BCS theory. Each item has a paragraph in italics which recalls its history, as rooted in the scientific biography of its main proponent. 
\begin{itemize}
\item Friedel. BCS picture with singularity in the density of  electronic states states.

This picture has been developed by Friedel and collaborators, based on their BCS theory in 1970 for some of the superconductors with the largest known superconductivity temperature known before 1986 (about 21 K, in the so-called $A15$ compounds). A basic assumption of this approach is that above the superconducting temperature  the metallic state is normal and has a well defined Fermi surface. The  interaction energy between electrons is supposed  in this picture to be much smaller than the band width, and electron-phonon interactions play the same decisive role as in the BCS theory. 

\textit{The Friedel proposal in 1970  had been based on the notion that in $A15$  compounds, the quasi one dimensional structure\footnote{A generic property of so called $A-15$ crystals such as $Nb_3Sn$ or $V_3Si$ .} may  lead to a large peak in the electronic density of states at the Fermi level. Within BCS theory, such a peak produces a larger superconducting temperature than a lower regular flat density of states.}

Friedel and collaborators argued that in CuO planes, such a singularity arises in the density of states precisely for one electron per lattice cell in the CuO lattice. Neglecting Coulomb interactions, they adapted their previous theory to the new material and attributed to phonons, as in the BCS theory, the role of glue between pairs of electrons. The only new experimental feature which this proposal takes into account is the crystal anisotropy, as shown in figure \ref{fig01}.

\item Schrieffer. The Spin  Bag picture.

 Schrieffer \cite{schrieffer1} developed a theory based on repulsive electron-electron interaction in a large energy band. The starting point is similar to Friedel's, namely the existence of a well defined Fermi Surface at temperatures larger than the superconducting temperature. In certain cases, a geometry of the Fermi surface (a so-called ``nesting property'') each electron spin will surround itself with  a cloud of spin polarization. Schrieffer argued that under certain conditions, two such clouds could favour the formation of electronic spin singlets. In other words, he suggested an alternative route to the superconducting phase: instead of an attraction between electrons mediated by phonons, as in the standard BCS phenomenon, the attractive electron-electron interaction is a result of spin fluctuations, which  are themselves the result of repulsive electron-electron interactions. 

\textit{Some twenty years before, Schrieffer and col.  had worked out  a spin fluctuation theory to account for the transport and thermal properties of almost magnetic metals such as $Pd$ and $Pd-Ni$ alloys \cite{schrieffer2,lederermills}. This theory argued that spin fluctuations destroy superconductivity.}

 On the face of the HTS discovery, a contradictory result was found to be possible, under conditions not explored previously!

\item Schulz. The Renormalization Group  with Van Hove singularity.

 An original picture was proposed by Schulz \cite{schulz}. He devised  a brilliant Renormalization Group (hereafter RG) approach to electron-electron interactions in a half filled square lattice. A Van Hove singularity is a singular property of the density of electronic states when the Fermi Surface and the boundaries of the Brillouin zone come in contact. This feature has important consequences. Schulz explored (before Schrieffer) the  new scenario  based on repulsive electron-electron interactions. Even though the starting point of his study is, as in item F and Schrieffer above, that of weak (compared to the band width energy $K$) electronic repulsive interactions $U$,  the RG treatment of the many electron-electron interaction processes is based on the recognition  that various different symmetry breaking phases, associated to different electronic processes, are competing, in the case of a square Fermi surface and Van Hove singularities\footnote{Which is the case in a half filled square lattice.}, with one another. Among them a superconducting phase with a novel order parameter dubbed d-wave is competing with an antiferromagnetic insulating one. Taking into account a slightly more realistic Fermi surface (not exactly a square) it was easy to show, using Schulz' approach, that the most stable phase could indeed be a d-wave superconducting one \cite{lmp}.

 \textit{Prior to the discovery of HTS, Schulz had been working on one dimensional electron systems, where the RG allows to sort out  the various (competing) symmetry breaking ground states. A surprising result of his RG approach, along the work in ref. \cite{lmp}, is that repulsive electron-electron interactions may lead to a stable superconducting phase. As for the  authors of ref.\cite{lmp}, they had been working, before the discovery of HTS in 1986 on realistic Fermi Surfaces in quasi 1-D conductors\footnote{Schulz' pioneering work is not always recognized. Due to his untimely death, Schulz, unfortunately, was not able to disseminate his results in conferences, meetings, etc.. }.}

\item Mott. The bipolaron picture.

 Mott \cite{mott} also attributed HTS to electron-ionic lattice interactions. Famous for his numerous contributions to physics, he  had -- among others -- published  pioneering studies of transition metal oxides, and of their electronic states.

\textit{ In particular he had pointed out, many years earlier, that due to strong electron lattice interactions in insulating  oxides, isolated electron would deform the lattice by Coulomb attraction between electrons and the ionic lattice, forming so-called polarons \cite{mott1}.} He then argued that in HTS  such polarons could form pairs (due to a lowering of elastic energy) dubbed bi-polarons,  which are bosons. Bosons then condense to form a superconducting state.

\item Anderson. The Resonating Valence Bond picture. The Mott insulator parent of HTS.

 P. W. Anderson (PWA) \cite{PWA} proposed a revolutionary ``Resonating Valence Bond'' (RVB) theory. His approach took the view, contrary to Friedel, Schrieffer and Schulz, that Coulomb interactions (of the order of $U=100 000 K$),  far exceed, in the CuO plane,  the band width energy (of the order of $K= 10 000 K$). His starting point was that the undoped cuprate was a Mott insulator. He first suggested that in the case of the square lattice with one electron per site, the ground state  could be  different from  the expected N\'eel AF state.   PWA instead proposed that neighbouring spins form singlets, and  described a massive entanglement of products of singlet states ``resonating'' with each other. At half filling, the ground state would be a new insulating  quantum spin liquid. But if a sufficient concentration of ``holes'' (empty states) are injected in the lattice, the doped system of entangled singlet states becomes superconducting with a temperature of order of the kinetic energy\footnote{This point was quickly corrected\cite{lederer2} to an estimate of the order of $K^2/4U \approx 250 K$.}. His initial formulation predicted a superconducting order parameter of symmetry s, as in the  BCS theory; this was quickly revised in favour of d-wave, for experimental and theoretical reasons. The d-wave singlet  is favoured by strong  repulsive Coulomb interactions, because it has zero amplitude for the two electrons of the singlet to be simultaneously present on the same site.

\textit{The RVB proposal had been first developed by PWA, in 1973 \footnote{In the Materials Research Bulletin.}, years before the advent of HTS,  as an alternative ground state to the N\'eel insulating  (AF) ground state for certain Mott insulators\cite{PWA2}. PWA gave an example of a ``railroad trestle'' (a model with two coupled infinite straight lines with a regular array of spin 1/2 electron and exchange coupling in-chain and inter-chain). This model provided a proof of the possibility of a RVB ground state with a lower energy than the conventional N\'eel one.}

His novel intuition in 1986 about the CuO square with doped holes was that the system of holes embedded in the RVB spin liquid would be superconducting. His original prediction of the magnitude of  the superconducting temperature  nurtured wild hopes for very high ones. Another novel feature of this proposal was the separation of spin and charge degrees of freedom into ``spinons'' and ``holons'', in the RVB state: a spectacular breakdown of the independent electron picture.

Anderson's proposal received enthusiastic support by a large fraction of theorists, because of its novelty and  its audacity which seemed to correspond to the novelty of the HTS material, and to that of spectacularly high  superconducting temperatures. Another sizeable part of the theorists' population, rejected it, precisely because of its novelty and audacity: attributing the new phenomenon to repulsive interactions, while from 1958 to 1986 all superconducting material had been understood on the basis of attractive (el-phonons) seemed hard to swallow...

\item Laughlin. The anyon superconductivity picture.

Within the new paradigm of a superconductivity scenario based on strong electron-electron repulsions, Laughlin \cite{laughlin} developped the so called ``anyon model of HTS'', also a daring  revolutionnary one. He focused also on the behaviour of electrons confined to a plane. Contrary to Anderson, he did not take into account the discreteness of the atomic lattice. His main new idea was that in the HTS phase,  a uniform magnetic field would spontaneously arise and organize electrons in a superfluid liquid similar to the Fractional Quantum Hall  liquid which he had so brilliantly studied in the early eighties and for which he was to be awarded the Nobel prize. In terms of audacity and novelty, his work passed that of Anderson.

\textit{ He used the wisdom acquired in the study of the Quantum Hall Effects (QHE), three years before the advent of HTS. He showed that a (2D) gas of anyonic particles had to be superconducting. Anyonic particles can only exist in 2D and have statistics governed by any phase angle upon interchange of two particles, contrary to bosons or fermions, the statistical angle of which is, respectively $2\pi$, or $\pi$. Anyons are believed to exist --on very firm theoretical grounds -- in the QHE.}

 A simple theoretical picture of the spontaneous emergence of a uniform magnetic field in a 2D lattice of strongly interacting electrons doped with holes was given in \cite{lpr} under the name ``flux phase''. A major novelty in Laughlin's scheme is that HTS might be the simultaneous spontaneous breaking of both gauge invariance and time reversal symmetries. 
\end{itemize}

The list of theoretical proposals given above is by no means complete, and is limited to the very early days of the scientific battles that are still developing today \cite{zaanen}. Qualitatively new viewpoints have emerged later on, which are not discussed in this paper. The aim of this paper is to reflect on  what we can learn  about the connection between thought and reality in the  science process from those few examples. I cannot guarantee here  historical rigor, which in this topic, as mentionned above, cannot be achieved by a single author. All I can say is that  I have tried to be as truthful as I could. 

\section{Underdetermination of theory by facts?}\label{underdetermination}

 At first sight, what I have sketched above seems to  support  the Duhem/Quine thesis on the underdetermination of theory by facts \cite{duhem,quine}.
All schemes listed above claim more or less to account for the same phenomena of HTS, on the basis of different hypothesis. 

As pointed out above, the differences between the theories of  the microscopic mechanisms of HTS have developed on the basis of universally accepted views, together with different ones on the relevant parameters for the cuprates. Within the various items I have listed above, there are  classes with basic differences, listed below, along with some overlap between those. 

\begin{itemize}
\item The band picture scheme.

 One class of proposals ( Friedel, Schrieffer, and Schulz for example) are based on the idea of the domination of kinetic energy $K$ over electronic interactions $U$. This is noted $K >> U$ and  is the so called band model, which assumes the band theory of metals to be  the correct starting point.

\item  The electron-phonon scheme.

  A sub class of the previous one  ( Friedel)  is based on the BCS model of electron-phonon interactions, and neglects electron-electron interactions; i.e. $U_{el-ph} > U_{el-el}$. Although I have only quoted Friedel in this context, a large body of theoretical work continues along that line.
	
\item  The doped Mott insulator scheme.

Items Schrieffer and Schulz represent another sub-class, which  takes the view that $K >> U_{el-el} >> U_{el-ph}$. Schrieffer and Schulz differ in that Schulz focuses on the Van Hove singularity. The latter is unavoidable for a half filled square lattice; Schulz seems to be the rigorous way to treat the Schrieffer ``spin bag'' idea, in the case of a square Fermi Surface touching the Brillouin zone.

\item The Mott insulator scheme.

 Another large class of theories is based on the notion that the undoped ground state of the  CuO planes is  a ``Mott insulator'', i.e. one where 
 $U_{el-el}>> K >> U_{el-ph}$. This is represented by items Anderson, Mott, Laughlin above.

\item  Class Mott  is simultaneous one that admits $U_{el-el} > K$  and that $U_{el-ph}$ plays a decisive role in the formation of  singlets (bi-polarons).

\item Items Friedel, Schrieffer, Schulz  contain the notion of a Fermi Surface in the normal (i.e. metallic) doped phase. Item M does not, since it relies on bosons which have no Fermi Surface\footnote{An interesting document which describes a partial state of affairs in the HTS battles in 1998 is listed in ref. \cite{bipolaron}.}. The existence of a Fermi surface is now an established experimental  fact.
\end{itemize}

In agreement with  experimental data, all theories share the view that HTS arises when the CuO planes are doped with holes\footnote{Inter CuO layer  coupling was considered later by PWA, then abandoned, due to  a crucial experiment. }.

An obvious lesson is that the development of theory is based on a choice of the relevant competing types of energy which dominate the HTS instability. This is, in my view, deeply rooted in the physicists' culture which teaches to analyse nature in most cases, in terms of competing energies, (i.e. ontological contradictions in the dialectical materialist sense) as discussed in \cite{lederer}. 

So is it relevant to appeal to the Duhem/Quine thesis as far as HTS is concerned? In spite of  appearances, I do not believe that; if at present there is underdetermination of theory by the facts, this cannot  be more than a temporary state of affairs. The Duhem/Quine thesis expresses a true fact...which has quite generally only a temporrary relevance, not a long lasting one. Sooner or later, scientific and/or technological progress allows to overcome this underdetermination. It took more than fifty years to establish  the universally accepted successful theory for low temperature superconductors.  For one thing there are numerous examples of phenomena the theory of which is undisputed, confirmed  by numerous experiments and verified predictions. For another, HTS erupted only thirty years ago, and no one may say that no  universally accepted  theory of HTS can ever  be reached in the future. Some of the theories listed above have been discarded on the basis of experiments, as discussed below. Others have had partial confirmation, but need better foundations and more experimental specific proof. Why should mankind be banned for ever from a completely satifactory theory of HTS? In a motionless world where the essence of things is eternal, the Duhem/Quine thesis may hold; not in our real historically evolving world.

What I have described in the last sections \ref{history}, and \ref{biography} seems to be a case study of what is discussed in section \ref{dialectics1}. Starting with the thought (theoretical activity) of various authors about the evolution (motion) of the electronic system in cuprates, from a materialist point of view, we oberve the contradictions of the evolving human thought about HTS. The thought (theories) about the objective transformation  of matter (emergence of HTS from non superconducting Copper oxides) exhibits the subjectivity of the motion of thought: each theoretical item in section \ref{biography} is narrowly linked with the specific scientific biography of its proponents. Each founder of an item listed in section \ref{biography} acts  both objectively and subjectively while picking subjectively from the material represented in figure \ref{structure} such objective features as connect to what he or she had successfully explained theoretically years before about an object, which, although different, also exhibits features similar to those of HTS  materials.

Another objective/subjective contradictory thought process is the HTS battle itself, which is waged through the scientific press and the various social forms (conferences, etc.) through which physicists -- theorists and experimentalists -- confront their views and their results, propose crucial experiments.  

  An increasing tendency  in this area is to form groups of searchers  convinced of the validity of a specific scheme, and to ignore contributions -- both theoretical and experimental ones -- which do not explicitly support it. International meetings are sometimes organized by groups who do not  invite contradictors...This might be the topic of a sociological study of the developpment of individual/social contradictions in the search for truth
in contemporary physics.
\section{Relativism?}

Should this lead to a relativistic attitude? Are theories, after all, as Kuhn wrote \cite{Kuhn}, beliefs shared by groups of persons? Obviously, if we consider the development of HTS theory from the start, this state of affairs reflects at least an appearance of truth. There are two objections to this : on one hand, it is a static statement which denies scientific evolution \cite{popper}. On the oher hand,  some of the initial theories have been discarded quickly, because some of their crucial predictions have been falsified by experiments. Laughlin's theory on HTS ref.\cite{laughlin,lpr} has failed  in its initial form because of a crucial experiment:  the predicted magnitudes of magnetic fields inside the material have not been observed. Similarly, the bi-polaron theory has been discarded by most, if not all, theorists. The initial prediction, both by Friedel's and Anderson's initial proposals, that the order parameter in HTS would have spherical ``s'' symmetry, has been quickly corrected :  the evidence that the order parameter in CuO HTS has ``d'' symmetry is overwhelming. The latter -- contrary with the situation in BCS supersonductores -- is unavoidable in case of a mechanism driven by el-el interactions. Those corrections, at firt sight, seem to support Popper's views \cite{popper2} on the progress of science. I have listed criticisms to Popper's falsification theory in ref.\cite{lederer}. Popper's views are  themselves refuted by the universal acceptance,,and widespread applications, of the superconductivity theory for simple superconducting metals. 

\section{Crucial experiment?}

Bacon \cite{bacon} has introduced the notion of crucial experiment, which allows to discriminate between a truthful theory and an erroneous one. Duhem \cite{duhem} has argued that there is no such thing. I have argued elsewhere \cite{lederer3} that Duhem's stand is destroyed by Duhem's own  admittance that `` the theory of vibrating strings is certain''. The notion of crucial experiment is vindicated by scientific practice and is intimately connected with the notion of truth, as I discussed in a previous paper \cite{lederer}.

 However, it is striking that, somewhat along some of Cartwright's ideas \cite{cartwright}, in the case of HTS, various different theories claim they account equally well -- sometimes equally badly -- for the same phenomena. Again this cannot be claimed as an everlasting state of affairs; more rationnally, it seems to point out that a theory incorporating all aspects of HTS needs to be established. The search for the  ``smoking gun'', in the HTS literature,  synonymous with ``crucial experiment'', is still going on. 

   Some experiments have been crucial to discard some theories. None, so far, has been able to pin point one and only one microscopic mechanism as the single correct one for HTS. 
	
	On the other hand, all theories discussed above have taken into account experimental features which are experimentally undisputable. What differs from author to author is the choice of the main objective contradictory features on which theory is constructed.
	
		It follows from this discussion that a first conclusion confirms, contrary to Kuhn, Cartwright, Rorty, Latour and others, that the process of knowledge is simultaneously subjective and objective, and reflects in contradictory social ways multiple ontological contradictions. 
	
	
	
	Science has known century long controversies, which have generally ended up in a qualitatively new theoretical scheme which supersedes previous disagreements, such as for example the debate about discreteness vs continutity for light. The historical process of research about HTS is by no means ended. Some now believe that, behind the present state of confusion,  astonishing theoretical progress on HTS  may be lurking around some laboratory corners. Far reaching consequences on our understanding of nature, far beyond the specifics of HTS itself  are hoped for \cite{zaanen}.
	
	\section{Dialectics of nature?}
	
	I have mentioned above the spontaneous tendency of physicists to think about natural processes in terms of conflicting scales of parameters and energies. 
	Under conditions such that one type of energy (or free energy, or thermodynamic potential)  parameter dominates, matter will express its essence through different phenomenal manifestations: order or disorder, insulator or conductor, ferromagnetism or paramagnetism, liquid or solid, symmetry or broken symmetry,  etc... Understanding the different relevant  scales of lengths, temperatures, magnetic fields, interactions, at work in the thing etc., is a basic skill of physicists to analyze nature. In the HTS problem, could it be that one specific aspect of the physics at hand is that various parameters are simultaneously relevant, while leading to conflicting types of order? It has  been understood  after a while that the primitive dichotomy between the various schools, based on whether $U_{el-el}$ is much larger, or  much smaller than  the band width $K$  is in fact not justified by  quantum chemical calculations. In fact, in the HTS cuprates, $U_{el-el} \approx K$. Under such conditions the role of $U_{el-ph}$, and its association/opposition to $U_{el-el}$ may be sometimes decisive for certain HTS phenomena. Other phenomena, not discussed in this paper, such as disorder, or the tendency of electrons to organize in stripes \cite{zaanen}, are believed to be of great importance. The spontaneous breaking of time inversion symmetry \`a la Laughlin  has not been supported by experiment in its original version, but may be revived when looking at finer details of  electronic structure. What is  at stake here is perhaps the relevance  or irrelevance of a theoretical approach  based on a binary contradiction of opposites.
	
	The ``pseudo gap'' phase seems to be an example where various order parameters compete with one another \cite{zaanen,laughlin2}. S\`eve \cite{seve} has discussed the extension of binary dialectics of nature to richer dialectics based on a multiplicity of poles. In technical words, the suggestion here  is that a number of free energy minima corresponding to different parameters, broken symmetries, etc., have similar magnitudes in the relevant temperature ranges, so that correlations within the material suffer multiple competing influences. This complexity seems also to be deeply connected with the structural and chemical complexity of doped HTS cuprates, in contrast with the simple structure of  metals of the BCS times. It may be that exploring this path, in the theoretical physics research, as in the philosophical one, is a way to a theory and to concepts which might supersede the present conflicts between the various chapels. 
	
	In ref.\cite{lederer}, in the discussion of Quantum Hall Effects (QHE), I have stressed that the physics of QHE is one where the energy scale $U_{el-el}$ is much larger than any kinetic energy scale. In that sense, the QHE physics is the opposite of simple metal physics, where the kinetic energy scale dominates. The reason for the lack of a present consensus about the HTS physics seems to be  connected to the intermediate situation, with a complexity calling for qualitatively new theoretical concepts, as well as new philosophical developments about the dialectics of nature.
	
	\section{Conclusion}
	
	The  initial development of the theoretical activity of physicists on the microscopic mechanism of HTS is a  rich story involving many aspects of the  practice and theory of knowledge. I have briefly pointed out some biographical and sociological features, but the main focus in this paper has been on the origins  of present day scientific pluralism. The story of HTS seems a specially well  suited  (experimental) ground for philosophical inferences about both the process of knowledge and those of  matter. Universal lessons have been proposed which are suggested, or confirmed, by this particular example. In particular, the materialist dialectic outlook on the motion of nature, and on the motion of thought about nature, which has proved its fruitfulness when bipolar contradictions are clearly at work, seems to need a deepening of its concepts when multipolar contradictions govern the 
	
	I apologize to the many physicists who have contributed significant results in the field of HTS, whom I have not cited, because of the focus of this paper, because of my own  subjective choices in describing the HTS early history, or simply because of ignorance.


\begin{thebibliography}{99}
\bibitem{seve}L. S\`eve, {\it{Sciences et dialectiques de la nature}}, La Dispute, Paris, 1998; {\it{Penser avec MArx aujourd'hui, Tome III `` La Philosophie?''}},La Dispute, 2014.
\bibitem{kojeve} Koj\`eve \textit{Introduction \`a la lecture de Hegel. Le\c cons sur la Ph\'enom\'enologie de l'esprit}  published  par Raymond Queneau. Paris, Gallimard, 1947.
\bibitem{aristo} Aristotle, {\it{ Metaphysica}}, ed. W. Jaeger, Oxford University Press 1957, eighth edition, 1985 
\bibitem{kant} Immanuel Kant {\it{  Kritik der reine Vernuft)}}, 1781.
\bibitem{hegel} G. W. F. Hegel {\it{ Wissenschaft der Logik}}, 1812–1816.
\bibitem{marx}K. Marx  {\it{Zur Krtitik de politischen \"{O}konomie}}, 1859.
\bibitem{engels1} Friederich Engels {\it{ Anti-During}} , 1878.
\bibitem{engels2} Friederich Engels {\it{ Ludwig Feuerbach und der Ausgang der klassischen deutschen Philosophie}}, 1886.
\bibitem{lenin}  V. I. Lenin, {\it{Materialism and Empirio-criticism}}, Intl Pub, 1970.
\bibitem{popper} K. Popper, \textit{Objective Knowledge, an Evolutionary Approach}, Clarendon Press, 1972.
\bibitem{dickson} M. Dickson, {\it{Plurality and Complementarity in Quantum Dynamics}}, Scientific Pluralism, S. H. Kellert, H. E. Longino and C. K. Waters, ed. Minnesota Studies in the Philosophy of Science, University of Minesota Press, 2006.
\bibitem{putnam} H. Putnam, {\it{Models and Reality}}, The Journal of Symbolic Logic, {\bf{45}}, 464, 1980.
\bibitem{cartwright} N. Cartwright, {\it{How the Laws of Physics Lie}}, Clarendon Paperbacks, 1983
\bibitem{hacking} I. Hacking, {\it{ Representing and Intervening. Introductory topics in the philosophy of natural science }}, Cambridge University Press, 1983.
\bibitem{Kuhn} T. Kuhn, {\it{ The Structure of Scientific Revolutions.}}, University of Chicago Press, 1962.
\bibitem{rorty} R. Rorty {\it{Objectivity, Relativism and Truth: Philosophical Papers I}}, Cambridge, Cambridge University Press, 1991.
\bibitem{latour} B. Latour, La Recherche, 01-03-1998-88927
\bibitem{lederer} Pascal Lederer, {\it{ The Quantum Hall Effects: a Philosophical Approach}}, Studies in the History and Philosophy of Modern Physics, {\bf{50}}(2015), 25-42.
\bibitem{chakra} B. K. Chakraverty and J. Ranninger, {\it{Bipolarons and superconductivity}}, Phil. Mag. Part B, {\bf{52 }}, 669, 1985.
\bibitem{bednorz} G.Bednorz and K. A. Mueller,{\it{  Z. Phys}}, {\bf{ B64}}, 189, 1986.
\bibitem{zaanen} J. Zaanen, arXiv: 1012.5461 v2 $\left[cond-mat.supr-con \right]$, 2011.
\bibitem{KO} H. Kamerlingh Onnes, {\it{ Leiden Com.}}, {\bf{120b, 122b, 124c}}, 1911.
\bibitem{MO} W. Meissner and R. Ochsenfeld, {\it{Naturwissenschaften }}, {\bf{21}}, 787, 1933.
\bibitem{BCS} J. Bardeen, L. N. Cooper and J. R. Schrieffer, Phys. Rev. {\it{108}}, 1175, 1957.
\bibitem{squid} see for example M. Tinkham, {\it{Introduction to Superconductivity}}, McGRAW-HILL International Editions, 1996;  B. D. Josephson, Phys. Lett., {\bf{1}}, 251, 1962.
\bibitem{degennes} P. G. de Gennes {\it{Les cristaux liquides n\'ematiques}}, Journal de Physique C5a, Suppl 10, {\bf{32}}, 1971.
\bibitem{duhem}P. Duhem, {\it{ Sauver les apparences. Sur la notion de th\'eorie physique de Platon \`a Galil\'ee }}, Vrin, {\it{Biblioth\`eque des Textes Philosophiques - Poche}}, 2004;  {\it{La th\'eorie physique}}, 1906, r\'e\'edition Vrin 1989.
\bibitem{quine}W. V. Quine, {\it{A System of Logic}}, Cambridge, Mass, Harvard, 1934.
\bibitem{friedel} J.Friedel, Physica C {\bf{153-155}}, 1610, 1988.
\bibitem{schrieffer1} J. R. Schrieffer, X.-G. Wen and S.-C. Zhang, Phys.Rev. Lett. {\bf{60}}, 944, 1988.
\bibitem{schrieffer2} N. F. Berk and  J. R. Schrieffer Phys. Rev. Lett. {\bf{17}}, 433, 1966.
\bibitem{lederermills} P. Lederer and D. L. Mills, Phys. Rev. {\bf{160}}, 590, 1968.
\bibitem{schulz} H. Schulz, Europhys. Lett. {\bf{4}}, 997, 1987.
\bibitem{lmp} P. Lederer, G. Montambaux, D. Poilblanc, J. Phys. France {\bf{48}}, 1613-1618 1987.
\bibitem{mott} N. F. Mott , Philos. Mag. B{\bf{58}}, 369, (1988);  J. Phys. France {\bf{50}}, 2811, 1989.
\bibitem{mott1} N. F. Mott, {\it{Metal-Insulator Transitions}}, Taylor \& Francis, London, 1974.
\bibitem{PWA} P. W. Anderson, G. Baskaran, Z. Zou, and T. Hsu, Phys. Rev. Lett. {\bf{58}}, 2790, 1967.
\bibitem{PWA2} P. W. Anderson,  Mater. Res. Bull. {\bf{8}}, 153, 1973.
\bibitem{lederer2}P. Lederer J. Phys. Soc. Jpn. {\bf{57}} 1729-1742, 1988.
\bibitem{laughlin} R. B. Laughlin, Phys. Rev. Lett. {\bf{50}}, 1395, 1983; Science {\bf{242}}, 525, 1988.
\bibitem{lpr} P. Lederer, D. Poilblanc and T. M. Rice, Phys. Rev. Lett., {\bf{63}}, 907, 1989.
\bibitem{althu} L. Althusser, {\it{Philosophie et philosophie spontan\'ee des savants}}, F. Maspero, 1974.
\bibitem{bipolaron} http://physicsworld.com/cws/article/news/1998/oct/02/superconductivity-debate-gets-ugly 1998.
\bibitem{bacon} Francis Bacon {\it{Novum Organum}} 1620.
\bibitem{popper2}K. Popper, {\it{Conjectures and refutations, the Growth of Scientific Knowledge.}}, London, Routledge and Kegan Paul, 1969
\bibitem{lederer3}  P. Lederer, {\it{Sur la th\`ese de Duhem sur la sous-d\'etermination}}, Cahiers rationalistes, {\bf 613}, 2011.
\bibitem{laughlin2} R. B. Laughlin, {\it{A different Universe: Reinventing Physics from the Bottom Down,}}, Basic Books, 2005.
\bibitem{seve-intro} L. S\`eve \textit{Une introduction à la philosophie marxiste}, 4;17, p. 301, Editions sociales, 1980.
\end{thebibliography}
\end{document}